\documentclass[conference]{IEEEtran}
\IEEEoverridecommandlockouts
\usepackage{cite}
\usepackage{amsmath}
\usepackage{amssymb}
\usepackage{amsfonts}
\usepackage{graphicx}
\usepackage{textcomp}
\usepackage{xcolor}
\usepackage{multirow}
\usepackage{soul}
\usepackage{tikz}
\usepackage{algorithm}
\usepackage{algpseudocode}
\usepackage{colortbl}
\usepackage{bm}
\usepackage{float}
\usepackage{filecontents}
\usepackage{CJK}
\usepackage{multirow}
\usepackage{algorithm}
\usepackage{algorithmicx}
\newcommand\encircle[1]{%
\tikz[baseline=(X.base)] 
  \node (X) [draw, scale=0.75, shape=circle, inner sep=0, fill=black, text=white, minimum size=0em] {\strut #1};}

\usepackage[a4paper, total={184mm,239mm}]{geometry}
\def\BibTeX{{\rm B\kern-.05em{\sc i\kern-.025em b}\kern-.08em
    T\kern-.1667em\lower.7ex\hbox{E}\kern-.125emX}}

\usepackage{todonotes}

\begin{document}

\title{\huge DRAM-Locker: A General-Purpose DRAM Protection Mechanism against Adversarial DNN Weight Attacks \vspace{-0.5em}
\thanks{This work is supported in part by the National Science Foundation under Grant No. 2228028, 2216772, and 2216773.}}

\author{Ranyang Zhou$^\dagger$, Sabbir Ahmed$^\ddagger$, Arman Roohi$^*$, Adnan Siraj Rakin$^\ddagger$, and Shaahin Angizi$^\dagger$ \\
\small $^\dagger$Department of Electrical and Computer Engineering, New Jersey Institute of Technology, Newark, NJ, USA\\
$^\ddagger$Department of Computer Science, State University of New York at Binghamton, NY, USA\\
$^*$School of Computing, University of Nebraska–Lincoln, Lincoln NE, USA\\ 
rz26@njit.edu, sahmed9@binghamton.edu, aroohi@unl.edu, arakin@binghamton.edu, shaahin.angizi@njit.edu \vspace{-2em}
\\}

\maketitle

\begin{abstract}
In this work, we propose DRAM-Locker as a robust general-purpose defense mechanism that can protect DRAM against various adversarial Deep Neural Network (DNN) weight attacks affecting data or page tables. DRAM-Locker harnesses the capabilities of in-DRAM swapping combined with a lock-table to prevent attackers from singling out specific DRAM rows to safeguard DNN's weight parameters. Our results indicate that DRAM-Locker can deliver a high level of protection downgrading the performance of targeted weight attacks to a random attack level. 
Furthermore, the proposed defense mechanism demonstrates no reduction in accuracy when applied to CIFAR-10 and CIFAR-100. Importantly, DRAM-Locker does not necessitate any software retraining or result in extra hardware burden. \vspace{-0.4em}
\end{abstract}


\section{Introduction}
\begin{figure}[t]
\begin{center}
 \begin{tabular}{ll}
 \begin{minipage}[c]{0.21\textwidth}
\includegraphics [width=1\linewidth]{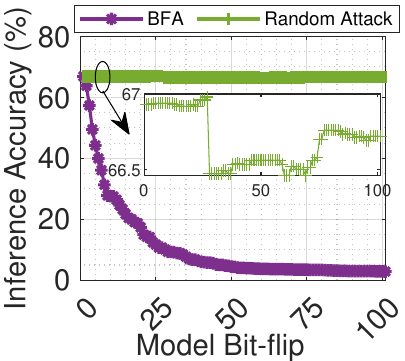} \end{minipage} &
 \begin{minipage}[c]{0.2\textwidth}
 \scalebox{0.76}{
\begin{tabular}{|l|c|}
\hline
DRAM Generation & $T_{RH}$  \\ \hline
DDR3 (old)      & 139K      \\ \hline
DDR3 (new)      & 22.4K     \\ \hline
DDR4 (old)      & 17.5K     \\ \hline
DDR4 (new)      & 10K       \\ \hline
LPDDR4 (old)    & 16.8K     \\ \hline
LPDDR4 (new)    & 4.8K - 9K \\ \hline
\end{tabular}
} 
 \end{minipage}
 \\ 
 \hspace{2.1 cm}   \small (a) & \hspace{2.1 cm}  \small (b)\\\vspace{-1em}
 \end{tabular} \vspace{-1em}
\caption{(a) Targeted bit flipping vs. random bit flipping for an 8-bit quantized VGG11 trained on CIFAR100, (b) RowHammer thresholds \cite{woo2022scalable}.} \vspace{-2em}
\label{RHthre}
\end{center}
\end{figure}

The widespread progress of Deep Neural Networks (DNN), achieving unparalleled performance and high accuracy even with models that have low bit-widths, has recently led to the emergence of various security-related attacks across many deep learning applications \cite{adi2018turning,rakin2019bit}. 
Recent research indicates that by identifying and manipulating a small set of susceptible bits within well-trained DNN weight parameters, adversaries can significantly degrade the resulting accuracy of these models as indicated in Fig. \ref{RHthre}(a) \cite{rakin2019bit}. These attacks, known as Bit-Flip Attacks (BFAs), have primarily been facilitated by a phenomenon in DRAM called RowHammer as the manifestation of a DRAM cell-to-cell interference and failure mechanism  \cite{kim2014flipping,mutlu2023fundamentally,kogler2022half}. On top of this, a novel class of adversarial fault injection techniques has been recently introduced \cite{saxena2023pt,zhang2020pthammer,frigo2020trrespass} that exploits BFA in memory addresses. Here an attacker can leverage the RowHammer attack to flip the bits in the page tables to corrupt the translation between the virtual and physical memory addresses. This gives attackers system-level privileges to overwrite a specific data block stored in a physical address using a replacement data block stored at a different physical memory address \cite{saxena2023pt}.

The RowHammer attack occurs when a malicious process repetitively activates and pre-charges a specific row (referred to as the aggressor row) until it reaches a certain threshold ($T_{RH}$) \cite{frigo2020trrespass,olgun2023dram}. This repeated activation induces bit-flips in adjacent rows (referred to as victim rows). Unfortunately, due to the shrinking size of DRAM chips in modern manufacturing processes, DRAM has become increasingly susceptible to RowHammer-induced bit-flips \cite{kim2020revisiting}. The data reported in Fig. \ref{RHthre}(b) shows a notable downward trajectory in the $T_{RH}$ over recent years. For instance, compared to DDR3 (new), LPDDR4 (new) requires approximately 4.5 times fewer hammering iterations to trigger the same effect \cite{woo2022scalable}. This trend suggests that $T_{RH}$ will nearly vanish with the advent of DDR5 \cite{marazzi2023rega}. 
To prevent RowHammer attacks, DRAM manufacturers and researchers have proposed hardware-based victim-focused defense mechanisms to proactively refresh the victim rows \cite{kim2014flipping,frigo2020trrespass}. However, such RowHammer mitigation proposals have faced a huge overhead both from latency and power consumption perspectives \cite{zhou2022lt}. To mitigate this, recent aggressor-focused swap-based mechanisms \cite{saileshwar2022randomized,woo2022scalable} proactively swap and unswap aggressors with random rows before reaching the $T_{RH}$. This raises another issue, i.e., the challenge of precisely monitoring the targeted row. SHADOW \cite{wi2023shadow} leverages unintelligent swap operations on all potential target rows to safeguard them from Rowhammer attacks. However, it is evident that attackers do not typically target numerous rows simultaneously, resulting in the wasteful execution of swap operations.

While there has been a multitude of generic victim-focused \cite{kim2014flipping,kim2014architectural} and aggressor-focused \cite{qureshi2022hydra,saileshwar2022randomized,woo2022scalable} mechanisms to protect the memory rows against RowHammer attacks, a few works have tried to tailor defense mechanisms for adversarial DNN weight attacks in software. Most of these methods try to make the model more robust by either using fewer bits for the model's weights~\cite{he2020defending} or by making the model larger to reduce the effect of weight noise on accuracy. Such mechanisms require software retraining or result in extra hardware burdens. 
In this work, we propose DRAM-Locker as a novel and advanced DRAM defense mechanism to protect the DNN against both adversarial BFA and page table attacks targeting weight parameters. Our contribution is summarized as follows.
\begin{itemize}
    \item We are the first to demonstrate a general-purpose DRAM defense mechanism with a lock-table that protects memory against DNN weight attacks affecting data or page tables;
    \item  We develop ISA, software support, and the interface required to implement in-DRAM swap operation; and
    \item  We extensively analyze the DRAM-Locker's applicability and efficiency in withstanding RowHammer vulnerability compared to recent hardware/software techniques over CIFAR-10 and CIFAR-100 DNN datasets.
\end{itemize}

\section{Background \& Motivation}
\subsection{DRAM}
\begin{figure}[t]
\begin{center}
\begin{tabular}{c}
\includegraphics [width=0.97\linewidth]{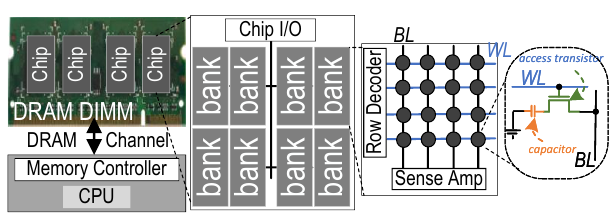}\vspace{-0.4em}
 \end{tabular} \vspace{-0.8em}
\caption{Organization of a DRAM chip.}\vspace{-1em}
\label{DRAM} \vspace{-1em}
\end{center}
\end{figure}

\vspace{-0.5em}
\textbf{Organization \& Commands.} The DRAM chip is a hierarchical structure consisting of several memory banks as shown in Fig. \ref{DRAM}. Each bank comprises 2D sub-arrays of memory bit-cells that are virtually ordered in memory matrices (mats), which have billions of DRAM cells on modern chips. Each DRAM bit-cell consists of a capacitor and an access transistor. The charge status of the bit cell's capacitor is used to represent binary ``1'' or ``0'' \cite{angizi2019redram,seshadri2017ambit,angizi2019graphide}. In idle mode, the memory controller turns off all enabled DRAM rows by sending the Precharge (PRE) command on the command bus. This will precharge the Bit-Line (BL) voltage to $\frac{V_{DD}}{2}$. In the active mode, the memory controller will send an Activate (ACT) command to the DRAM module to activate the Word-Line (WL). Then, all DRAM cells connected to the WL share their charges with the corresponding BL. Through this process, BL voltage deviates from the precharged $\frac{V_{DD}}{2}$. The sense amplifier then senses this deviation and amplifies it to $V_{DD}$ or 0 in the row buffer. The memory controller can then send read (RD)/write (WR) commands to transfer data to/from the sense amplifier array \cite{zhang2023aligner}.

\textbf{RowClone.}
Taking advantage of the DRAM's ability to transfer a complete data row to the corresponding row buffer during read operation, RowClone \cite{seshadri2013rowclone} has been devised as an uncomplicated and highly effective technique for facilitating a rapid in-memory copy operation (completed in less than 100 nanoseconds) within the DRAM sub-array, allowing data to be transferred from a source row to a destination row. RowClone eliminates the need to transfer data over the memory channel. The memory controller manages this by issuing two back-to-back ACT commands first to the source and then the destination without PRE command in between with almost negligible cost. By using this method, the latency and power consumption of a bulk copy operation can be reduced by a factor of 11.6 and 74.4, respectively \cite{seshadri2013rowclone}.

\subsection{SOTA Defense Mechanisms}
Multiple software and hardware mitigation mechanisms have been proposed to reduce the impact of RowHammer-based attacks \cite{frigo2020trrespass,lee2019twice,zhou2022red,zhou2023dnn,zhou2023p}. The standard mitigation approach used by manufacturers such as Apple \cite{Apple} is to reduce the refresh period, e.g., from 64ms to 32ms. 
The system manufacturers tend to increase refresh rates and hardware RHP \cite{kim2020revisiting}. Along this line, Target Row Refresh (TRR) \cite{frigo2020trrespass} and other counter-based detection methods \cite{qureshi2022hydra,seyedzadeh2016counter} require add-on hardware to calculate rows' activation and record it to other fast-read-memory (SRAM \cite{lee2019twice}/CAM \cite{park2020graphene}). The controller will then refresh the target row if the number reaches $T_{RH}$ \cite{frigo2020trrespass}. However, such proactive refreshing proposals have faced a huge overhead both from latency and power consumption perspectives. A method called Secure Row-Swap (SRS) \cite{woo2022scalable} has demonstrated the use of fewer counters for crucial data and implemented associated threat mitigation using the swap operation. However, such mitigations not only impose a significant slowdown to the system but also require a pre-defined threshold at CPU design time. To solve this issue, LT-PIM protects only a critical part of the memory rather than the complete data \cite{zhou2022lt}. Mitigations of this kind are susceptible to breakthrough attacks on many bits, such as Half-Double \cite{kogler2022half}, which take advantage of previously unknown access patterns or Threshold Breaker \cite{zhou2023threshold}.  
Software-only schemes offer the advantage of compatibility with existing hardware, making them more deployable. However, it's important to note that existing software-only mitigations require modifications to the memory allocator and may not be fully effective. The software-based efforts such as SoftTRR \cite{zhang2022softtrr} enable software tracking of activations to Page Table Entry (PTE) rows and accordingly issue mitigation. Similar to hardware-based TRR, the efficacy of such defense reduces when the attack happens on more bits and larger distances as in Half-Double \cite{kogler2022half}. CTA \cite{wu2019protecting} is susceptible to privilege escalation attacks such as PThammer \cite{zhang2020pthammer} targeting L1PTE. There are a few defense mechanisms developed especially for exploits on page tables. SecWalk \cite{schilling2021secwalk} can detect only up to 4-bit flips per PTE with error detection codes though it is susceptible to attacks like ECCploit \cite{cojocar2019exploiting}. PT-Guard \cite{saxena2023pt} presents a method that tracks the rows under attack by asserting the Message Authentication Code (MAC) to the PTE. In DRAM writes, they split the MAC into eight parts and embed them in the unused bits of the PTE. Then they recomputed the MAC and performed integrity checks on the hardware page table to ensure the PTE remained unchanged. In our work, there is no need to occupy the resources of software sides without reducing the performance.

\section{Threat Model}
The DRAM-Locker is designed to safeguard DRAM against the following adversarial DNN weight attacks.

\begin{figure}[t]
\begin{center}
\begin{tabular}{c}
\includegraphics [width=1.03\linewidth]{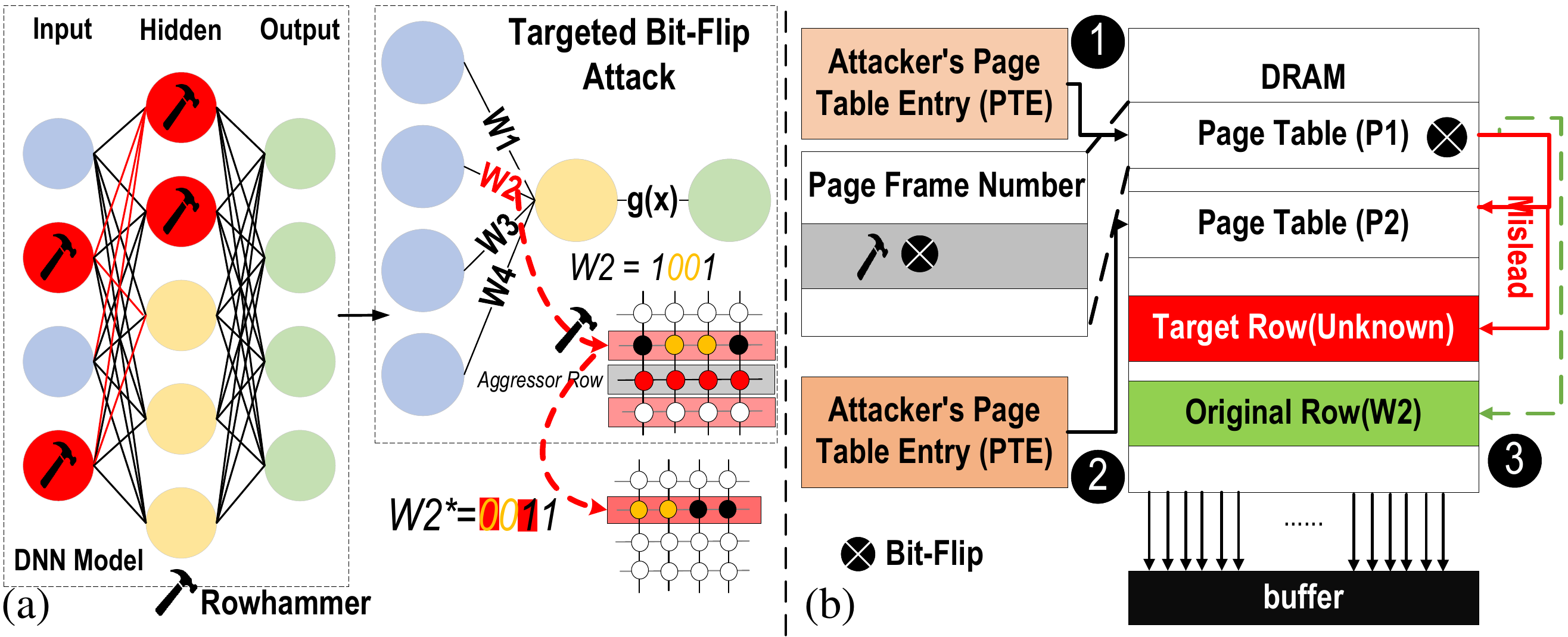}\vspace{-0.4em}
 \end{tabular} \vspace{-0.8em}
\caption{Two major DNN threat models in this work: (a) BFA, (b) PTA.}\vspace{-1em}
\label{ptattack} \vspace{-1.1em}
\end{center}
\end{figure}

\textbf{Bit-Flip Attack (BFA).}
We assume the following threat model for BFA. 1) DNN model inference is running on a resource-sharing environment which is practical due to the recent popularity of Machine-Learing-as-a-Service (MLaaS)~\cite{ribeiro2015mlaas}. The attacker can run user-level un-privileged processes remotely on the same machine deployed by the victim DNN model;
2) Each DRAM row has a threshold $T_{RH}$ after becoming an aggressor row, and once exceeded within the refresh interval ($T_{ref}$), it will impose a bit-flip to two adjacent victim rows as depicted in Fig. \ref{ptattack}(a);
3) We assume that all vulnerable data rows are neither concentrated in one/two sub-arrays nor evenly distributed in each sub-array. Experimentally, most sub-arrays store several data rows simultaneously; some may store multiple or none;  
4) The attacker has a detailed mapping file that can locate the physical address of the target data in the neural network and is aware of the initial static mapping of the DRAM rows (i.e., physical adjacency information between rows) 
\cite{wi2023shadow};
and 5) the attacker is aware of the internal structure of the DNN models, e.g., the number of layers and the width of each layer. On top of that, the attacker has complete knowledge of the DNN model parameters, their values, and bit representation for inference. They can catch the bits that can mostly reduce the accuracy, such as W2 shown in Fig. \ref{ptattack}(a).

\textbf{Page Table Attack (PTA).}
We assume the following practical threat model for PTA established by prior works \cite{saxena2023pt}. 1) The attacker can map the virtual addresses to physical addresses using several techniques such as leveraging huge page support, hardware-based side-channel attack~\cite{gruss2018another}, and memory messaging~\cite{kwong2020rambleed};
2) The attacker can cause a targeted bit-flip to the page table and cause a bit-flip at the desired location using fast and precise multi-bit-flip techniques~\cite{yao2020deephammer}; 3) We assume the kernel and operating system are trusted and well-protected;
4) Following standard practice, we assume the commercial DRAM is not protected by ECC and cannot protect large-scale deep learning models against RowHammer~\cite{yao2020deephammer}; and
5) Attacker can insert their own virtual page table entry, as depicted in Fig. \ref{ptattack}(b)\encircle{1} and \encircle{2}  perform a bit-flip within their page table entries P1, coercing them to direct to a secondary virtual address row P2 \cite{saxena2023pt}. Knowing the location of the target row after flipping the page table, the attacker can redirect the pointer to the target row out of range.

\section{DRAM-Locker}
The DRAM-Locker aims to serve as a comprehensive remedy for various forms of severe RowHammer attacks targeting DNN weight parameters. The core idea behind DRAM-Locker is to prevent RowHammer attackers from singling out specific DRAM rows by securing those rows with locks. Subsequently, DRAM-Locker employs the SWAP command to interchange the locked row with a free one, effectively unlocking it and restoring normal program execution. Compared with the previous counter-based designs \cite{frigo2020trrespass,seyedzadeh2016counter}, DRAM-Locker only requires a fast read-and-write memory space for the lock-table without any counter overhead. The DRAM-Locker also offers scalability and flexibility w.r.t. various applications. In other words, our framework allows users to customize the data they are willing to protect without requiring changes to the framework. As compared with other swap-based frameworks such as \cite{wi2023shadow}, DRAM-Locker is more directional and can effectively reduce the overhead of useless and time-consuming protection.

\subsection{Protection Framework}
Our protection framework exploits the lock-table to record the physical addresses of the DRAM that need to be locked. These addresses depend on the physical address of the data the user is willing to protect. For example, frequently used data is often the target of attackers, so adjacent rows of these data can be recorded in a lock-table. In addition, due to the uncertainty of attacks, users can manually add any row that has a high probability of becoming an aggressor row into the the lock-table. As depicted in Fig. \ref{ptattack}(a), conventional BFA techniques have been proven to be highly inefficient when applied to large-scale DNNs with extensive datasets. They often perform a large number of bit-flips on weights and/or activations without effectively reducing the accuracy of the model. But if the attacker focuses on attacking adjacent rows of particular weights to cause bit-flip, this will cause fatal damage to the DNN model.
We propose to segregate and retain these addresses within a lock-table in SRAM. Once the addresses are stored in the lock-table, any attempt to access this data without the accompanying unlock command will result in a block. In DNNs, weights typically represent high-frequently used data. Locking these weights can lead to substantial delays and a significant burden on the controller. Additionally, frequent access implies continuous refreshing, and the possibility of bit-flips occurrence within these rows is definitely low. If we choose to lock frequently used rows, then we have to unlock them constantly, so locking adjacent rows can substantially decrease the need for frequent unlocks.  
 \begin{figure}[t]
\begin{center}
\begin{tabular}{c}
\includegraphics [width=0.98\linewidth]{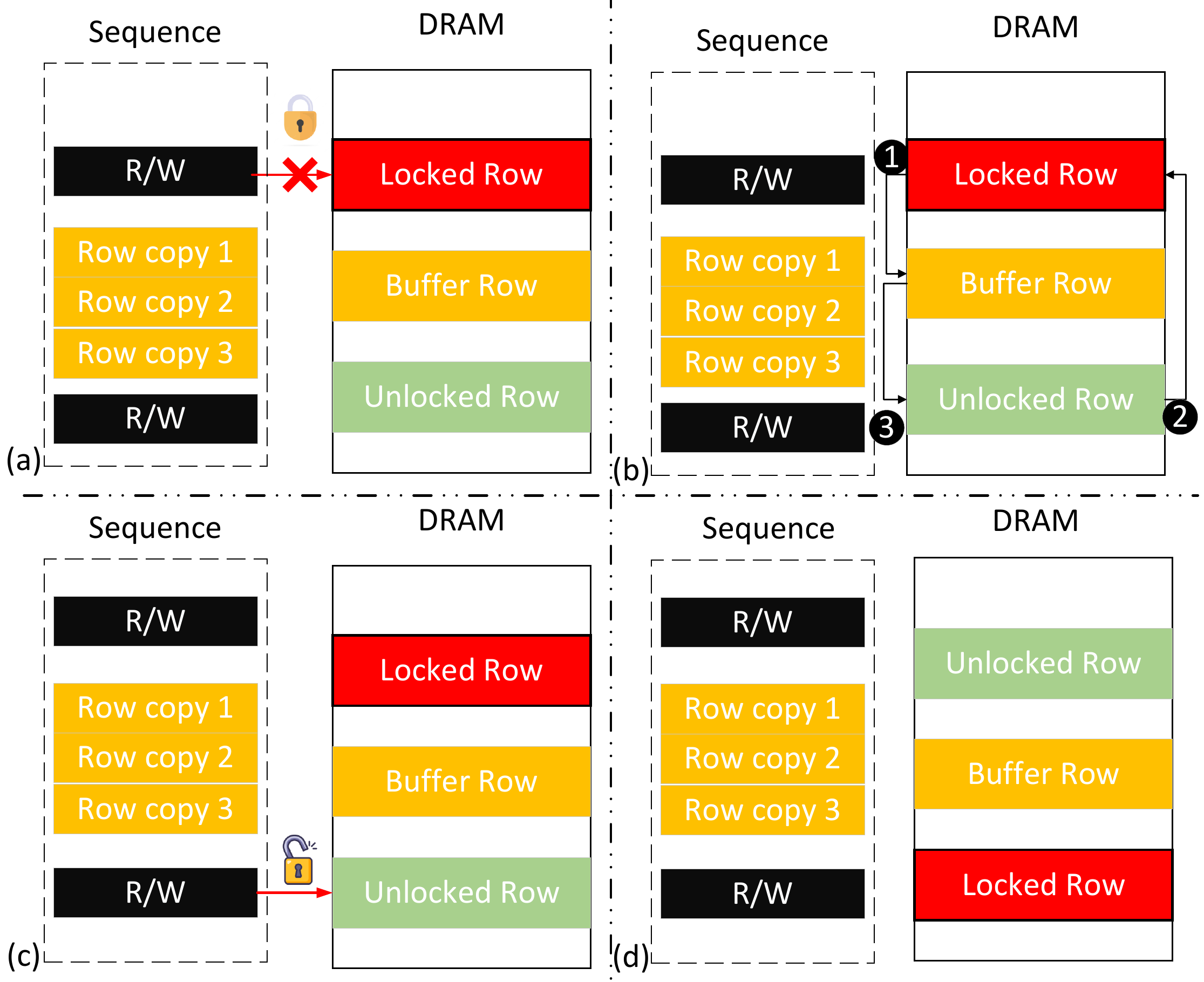}\vspace{-0.4em}
 \end{tabular} \vspace{-0.8em}
\caption{Overview of the DRAM-Locker. (a) R/W of the Locked Row, (b) Implementation of a SWAP operation, (c) R/W of the Unlocked Row, (d) Final status after updating the lock-table.}
\label{DL} \vspace{-2.2em}
\end{center}
\end{figure}
 
DRAM-Locker prepares for subsequent lock recognition and unlocking operations by storing instructions in Sequence. As shown in Fig. \ref{DL}(a), if an R/W (Read/Write) instruction in the Sequence contains the locked address, access to the Locked Row will be denied. Therefore, the DRAM-Locker needs SWAP operations to unlock. As shown in Fig. \ref{DL}(b), SWAP operations are transferred into DRAM instructions via ISA. The DRAM implements the sets of Row Copy instructions in order to perform the SWAP operation to pull out the Locked Row. Besides, since the RowClone principle is to overwrite the data of the copied row into a new row, we devise a Buffer Row. The detailed steps are as follows. In step \encircle{1}, DRAM-Locker copies the Locked Row to the Buffer Row; in step \encircle{2}, it copies the Unlocked Row to the Locked Row; and in step \encircle{3}, DRAM-Locker copies the Buffer Row to the Unlocked Row. This step does not change the lock-table, but exchanges the original data of the two addresses, ultimately achieving the unlocking process. The data stored at the locked address has been swapped into an unlocked row. Fig. \ref{DL}(c) shows that the next R/W instruction requires the address of the Unlocked Row to access the data of the original Locked Row. Taking the $T_{RH}$ into account, it becomes necessary to re-secure the previously unlocked rows by periodically refreshing the lock-table. In Fig. \ref{DL}(d), for instance, following the completion of a SWAP operation, the controller reinstates the swapped address into the lock-table after a cumulative count of 1k R/W instructions, effectively re-locking the data row.

\subsection{Lock-table \& Sequencing}
The lock-table plays a crucial role in our platform similar to that of the count-table in state-of-the-art designs \cite{lee2019twice, seyedzadeh2016counter}. Meanwhile, lock-table has no dependency on extra components such as counters. While counter-based designs employ a count table to monitor memory access patterns and keep track of the number of accesses to each row in the memory, DRAM-Locker stores the addresses of vulnerable rows and denies permission-less access requests to them. In counter-based designs, when a row is accessed excessively within a short time frame, the count-table is the indicator of a potential RowHammer attack. In response, the defense can take preventive actions to mitigate it. In DRAM-Locker, whenever the controller recognizes an R/W instruction, it will traverse the lock-table. If the address contained in the instruction is found, the instruction will be skipped. Therefore, no matter how many requests the attacker sends, they will be invalid and the instructions will not be executed. As a result, the latency caused by invalid instructions will also be eliminated. The instructions from attackers will also be stored in Sequence. If these instructions are skipped, we can also improve the performance.
When the program necessitates accessing these data rows, the SWAP operation is required to unlock the secured rows. So we insert row copy instructions to the sequence in order to execute the SWAP operations. To realize this, DRAM-Locker requires ISA support which will be introduced in the next subsection. DRAM-Locker uses three Row Copy instructions to implement the SWAP operation.

\subsection{ISA Support}
\begin{figure}[t]
\begin{center}
\begin{tabular}{c}
\includegraphics [width=0.96\linewidth]{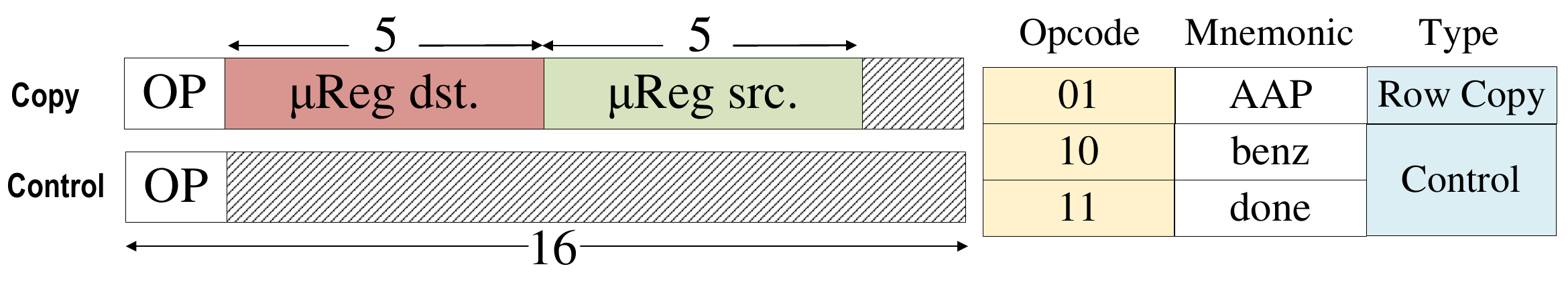}\vspace{-0.4em}
 \end{tabular} \vspace{-0.8em}
\caption{ DRAM-Locker’s instructions, $\mu$Ops, and their description.}
\label{ISA} \vspace{-1.6em}
\end{center}
\end{figure}

The DRAM-Locker is intentionally crafted as a general-purpose autonomous defense mechanism. Consequently, it needs to be accessible to system-level libraries and developers. From a developer's perspective, DRAM-Locker needs to transfer the SWAP operations and insert some specific instructions to perform the defense. Therefore, to enable widespread RowHammer defense for both BFA and PTA, ISA support is indispensable for any user-level program. This can be translated into the DRAM-Locker's hardware instruction set during installation.
As shown in Fig. \ref{ISA}, DRAM-Locker is designed to process two 16-bit instruction types after compiling the upper-level code: (1) a copy instruction based on RowClone method \cite{seshadri2013rowclone}, (2) an instruction for control operations. When OP is 01, DRAM-Locker performs a row copy operation by activating both $\mu$Reg src. and $\mu$Reg des. Opcodes 10 and 11 correspondingly denote straightforward control operations for managing loops and termination within the DRAM-Locker's control flow., i.e., {\tt bnez}, {\tt done}.

\subsection{Discussion: Challenges with Unsuccessful Swapping.}
We conducted comprehensive circuit-level simulations to investigate the impact of process variations of in-DRAM SWAP as the key operation in the DRAM-Locker based on the framework discussed in Section V. Our study considered a worst-case DRAM cell scenario where all components (including cell/BL/WL capacitance and transistor in Fig. \ref{DRAM}) exhibited variation. We performed a Monte Carlo simulation using the 45nm NCSU PDK library. This simulation comprised 10,000 trials, and we systematically increased the variation in parameters from $\pm$0\% to $\pm$20\%. It is worth noting that reducing the transistor size is anticipated to exacerbate the impact of process variation. Our observation shows that the percentage of erroneous SWAP operation across 10,000 trials is 0\%, 0.14\%, and 9.6\% for $\pm$0\%, $\pm$10\%, and $\pm$20\%, respectively.

\section{Performance Evaluation}
\begin{figure}[h]
\begin{center}
\begin{tabular}{c}
\includegraphics [width=0.82\linewidth]{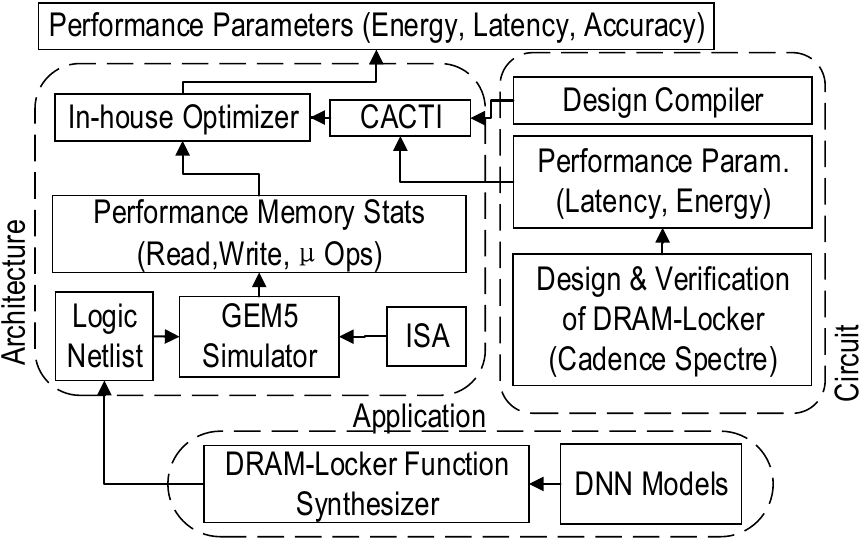}\vspace{-0.4em}
 \end{tabular} \vspace{-0.2em}
\caption{Proposed cross-layer evaluation framework.}
\label{flowchart} \vspace{-1.2em}
\end{center}
\end{figure}

\textbf{Setup \& Framework.}
We present a cross-layer evaluation framework as depicted in Fig. \ref{flowchart} to demonstrate the benefits of DRAM-Locker in protecting memory against adversarial DNN weight attacks. Firstly, we develop DRAM-Locker's sub-arrays with peripherals using Cadence Spectre in the 45nm NCSU PDK library \cite{NCSU_PDK} at the circuit level to verify SWAP functionality, attain performance parameters, and measure the row-shuffle time. The memory controller and registers are designed and synthesized by Design Compiler with a 45nm industry library. Afterward, we incorporated the results from circuit-level assessments and extensively modified CACTI at the architecture level. Next, we implemented DRAM-Locker's ISA in gem5 \cite{binkert2011gem5}, and exported the memory statistics and performance to an in-house C++ DRAM-Locker optimizer, taking the CACTI output and application netlist as the inputs. At the application, we evaluated the performance of our proposed technique in defending against adversarial BFA and PTA using various DNN models and datasets, where the weights are quantized to 8-bit width. To carry out the BFA and PTA, we randomly sampled images from the test/validation set, with a default sample size of 128 for both datasets.

\textbf{Hardware Overhead Analysis.}
We compare the DRAM-Locker’s hardware overhead with the latest RowHammer mitigation mechanisms in the literature in Table \ref{eva}. In this experiment, we utilize a uniform configuration of 32GB:16-bank DDR4 DRAM for all frameworks. The aim is to standardize capacity and area overheads across different frameworks. In Table \ref{eva}, {i)} the involved memory refers to the type of memory utilized by the framework for RowHammer protection. As previously discussed, certain frameworks rely on counters to monitor intrusions and store tracking information within the system, utilizing Content-Addressable Memory (CAM) or Static Random-Access Memory (SRAM). However, CAM and SRAM are significantly more costly in comparison to DRAM with the same overhead. Therefore, opting for a framework with such additional resources may lead to debates regarding cost-effectiveness. Take the Graphene \cite{park2020graphene} as an example, it employs both CAM and SRAM and the total capacity overhead is 0.53MB + 1.12 MB. It is worth pointing out that while most of the frameworks' area overhead in Table \ref{eva} is already quite compact, in terms of practical circuit design, minimizing alterations to the existing structure is generally more pragmatic. We observe that SHADOW \cite{wi2023shadow} and DRAM-Locker promote the use of extremely less extra components for constructing their defenses. Therefore, these two frameworks are selected for further security and performance analysis in the next part.

\begin{table}[b] \vspace{-1em}
\caption{Comparison with prior RowHammar mitigation frameworks.} 
\begin{center}\vspace{-1em}
\scalebox{0.85} {
\begin{tabular}{ccccc}
\hline
Framework                                                  & involved memory & capacity overhead & area overhead  \\ \hline
Graphene \cite{park2020graphene}          & CAM-SRAM        & 0.53MB$^\ddagger$+1.12MB$^\dagger$      & 1 counter      \\
Hydra \cite{qureshi2022hydra}             & SRAM-DRAM       & 56KB$^\dagger$+4MB$^*$          & 1 counter      \\
TWiCE \cite{lee2019twice}                 & SRAM-CAM        & 3.16MB$^\dagger$+1.6MB$^\ddagger$      & 1 counter      \\
Counter per Row                                            & DRAM            & 32MB$^*$              & 16384 counters \\
Counter Tree \cite{seyedzadeh2016counter} & DRAM            & 2MB$^*$               & 1024 counters  \\
RRS \cite{saileshwar2022randomized}                & DRAM-SRAM      & 4MB$^*$+NR$^\dagger$          & NULL           \\
SRS \cite{woo2022scalable}                & DRAM-SRAM      & 1.26MB$^*$+NR$^\dagger$        & NULL           \\
SHADOW \cite{wi2023shadow}                & DRAM            & 0.16MB$^*$            & 0.6\%          \\
P-PIM \cite{zhou2023p}                 & DRAM            & 4.125MB$^*$           & 0.34\%         \\
\emph{DRAM-Locker}                        & \emph{DRAM-SRAM}      & \emph{0+56KB$^\dagger$}          & 0.02\% \\ \hline
\end{tabular}}
\end{center} 
\label{eva}

\tiny NR = Not Reported\\
$^*$\tiny The capacity overhead of DRAM.
$^\dagger$\tiny The capacity overhead of SRAM. 
$^\ddagger$\tiny The capacity overhead of CAM. \vspace{-3.4em}
\end{table}

\begin{figure}[t]
\begin{center}
\begin{tabular}{c}
\includegraphics [width=1.01\linewidth]{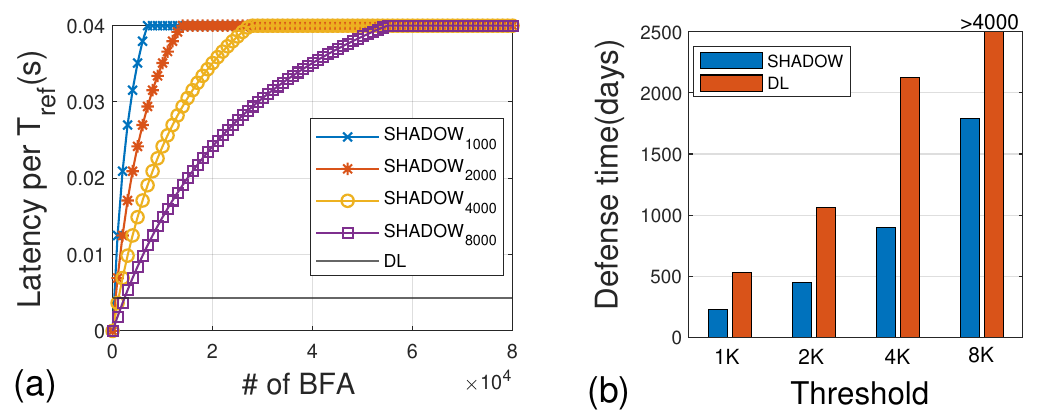}\vspace{-0.4em}
 \end{tabular} \vspace{-1.2em}
\caption{(a) Latency of DRAM-Locker (indicated by DL) and SHADOW \cite{wi2023shadow} in different
numbers of BFA, (b) Defense time (per day) in various thresholds.}
\label{security} \vspace{-1.9em}
\end{center}
\end{figure}

\textbf{Security \& Performance Analysis.} 
Figure \ref{security}(a) shows the comparison between latencies of  SHADOW \cite{wi2023shadow} with $T_{RH}$=1k, 2k, 4k, and 8k, and the DRAM-Locker with $T_{RH}$=1k. The threshold here represents the number of necessary visits to the aggressor row by the attacker to ensure that the victim row can generate a bit-flip. Whether in PTA or BFA, the attacker must ensure that the bit they expect is flipped accurately, so in the experiment, we consider the threshold as the minimum number of times. Also in this experiment, we assume that DRAM-Locker suffers from a 10\% error rate due to unsuccessful SWAPs of instructions. Please note that if the DRAM-Locker shows no errors then it will be ideally invulnerable.
In Fig. \ref{security}(a), due to the DRAM-Locker's distinctive SWAP mechanism, we exclusively evaluate the worst case, which is the 1k threshold. In this case, our framework demands the highest number of SWAP operations, resulting in the longest latency.
We can see that SHADOW \cite{wi2023shadow} exhibits a defense threshold. System integrity is compromised once this threshold is surpassed, signifying that further delay escalation is halted. Conversely, DRAM-Locker operates without such a defense threshold. Hence, the latency in DRAM-Locker is solely contingent on $T_{RH}$, with the worst case in our model being a $T_{RH}$=1k. Unlike SHADOW, our framework not only boasts remarkably low latency but also demonstrates general applicability across various DRAM chips, without any distinctions. In Fig. \ref{security}(b), we conducted experiments to evaluate the duration for which SHADOW and DRAM-Locker can effectively defend against attacks. Considering a 10\% error rate during the execution of Row Copy instructions, we take the expected outcomes of these error instructions into account. Then, we deemed success as achieving a probability exceeding 99\%. This means in both PTA and BFA threat models, when the attacker aims to flip the target bit, the probability is lower than 1\%. As a result, we observe that even considering a 10\% error to each row copy, DRAM-Locker could maintain an effective defense mechanism against attacks for a period exceeding 500 days under the 1K threshold. Therefore, in contrast to SHADOW, our framework exhibits reduced latency and extended defense duration.

\begin{figure}[t]
\begin{center}
\begin{tabular}{c}
\includegraphics [width=0.98\linewidth]{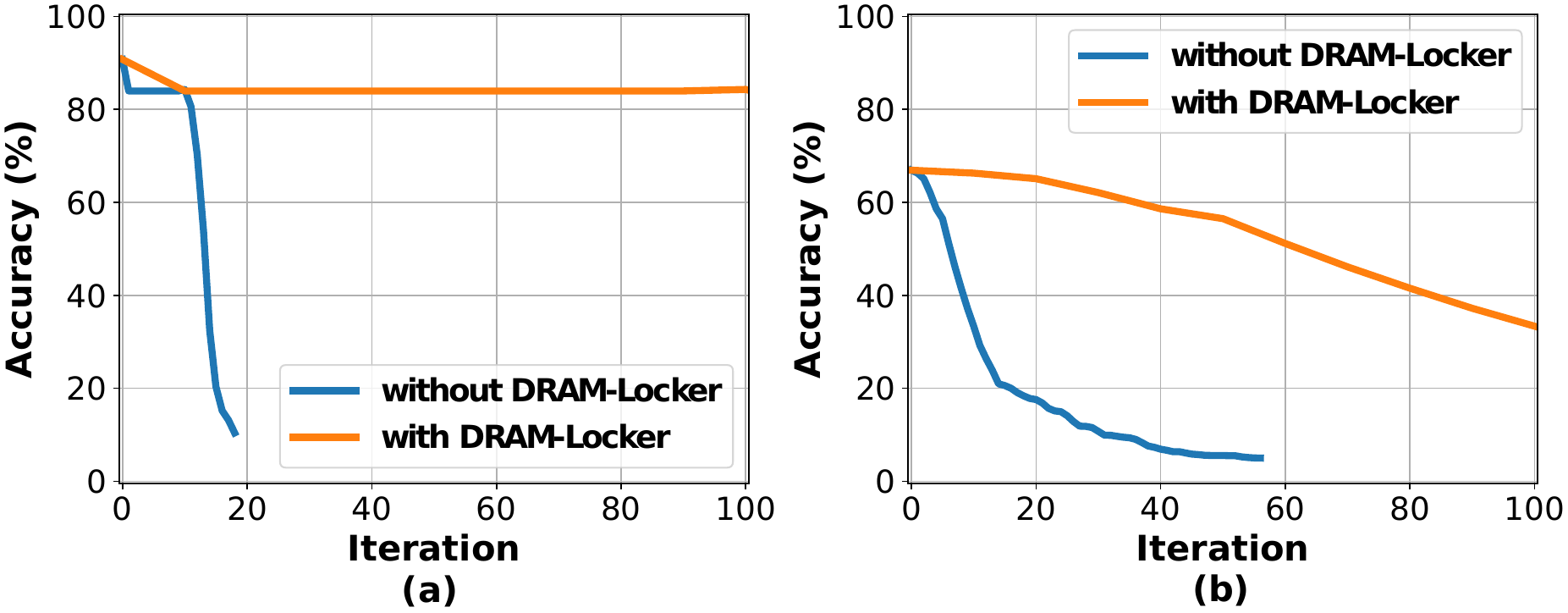}\vspace{-1.0em}
 \end{tabular} \vspace{-0.2em}
\caption{Proposed DRAM-Locker evaluation for (a) ResNet-20 trained on CIFAR-10, (b) VGG-11 trained on CIFAR-100. The degradation in performance is shown for 100 iterations of the attack. Incorporating DRAM-Locker makes it increasingly difficult for the attacker to cause performance degradation.}
\vspace{-1.0em}
\label{fig:cifar-results}
\end{center}
\end{figure}

\noindent\textbf{Evaluation of DRAM-Locker against BFA \& PTA.} Figure~\ref{fig:cifar-results} shows the efficacy of DRAM-Locker in alleviating the performance degradation caused by BFA. Note that this evaluation considers the worst case $\pm$20$\%$ variation on all DRAM components and thus considers that the BFA is successful 9.6$\%$ of the time as discussed before. The figure presents performance degradation across two different evaluation models, ResNet-20 trained on the CIFAR-10 dataset and VGG-11 trained on the CIFAR-100 dataset. The plots clearly reveal that with DRAM-Locker, it takes the attacker an increasing number of iterations to cause the same performance degradation. Thus, DRAM-Locker significantly increases the computational overhead for the attacker by necessitating more bit-flips to achieve equivalent attack effectiveness. We ran the same experiment considering PTA as the attack mechanism. The findings reveal that the attacker similarly needs a growing number of iterations to induce an equivalent performance decline.

\textbf{Comparison to other Defenses.}
In Table~\ref{tab:cmp}, we compare our proposed DRAM-Locker, with existing training-based defenses~\cite{rakin2021ra,he2020defending,li2020defending}. Most of these methods try to make the model more robust by either using fewer bits for the model's weights~\cite{he2020defending} or by making the model bigger to reduce the effect of weight noise on accuracy~\cite{rakin2021ra}. When we use DRAM-Locker to protect a specific number of vulnerable bits (for example, 1150), it does a better job of defending the BFA than the binary model. However, it's worth noting that our method does add a small amount of delay and energy use, which isn't an issue for the training-based defenses~\cite{rakin2021ra,he2020defending,li2020defending}. On the other hand, the existing training-based methods often come with downsides like taking a lot of time to train and reducing the model's accuracy. Our DRAM-Locker doesn't have these problems; it works well against BFAs and PTAs without needing extra training time or hurting performance, and only adds a minor hardware cost. Plus, our method can be used alongside existing software protections or training-based defenses~\cite{he2020defending,rakin2021ra} to make the model even more secure against various types of attacks.

\begin{table}[h]
\centering
\caption{{Comparison to other competing software defense methods on CIFAR-10 dataset evaluated attacking a ResNet-20 model.}}
\label{tab:cmp}
\scalebox{0.8}{
\begin{tabular}{|c|c|c|c|}
\hline
\begin{tabular}[c]{@{}c@{}} {Models} \end{tabular} & \begin{tabular}[c]{@{}c@{}}{Clean Acc.}(\%)\end{tabular} & \begin{tabular}[c]{@{}c@{}}{Post-Attack acc.}(\%)\end{tabular} & \begin{tabular}[c]{@{}c@{}} {Bit-Flips \#} \end{tabular} \\ \hline
Baseline ResNet-20~\cite{rakin2019bit}  & 91.71 & 10.90 & 20 \\
Piece-wise Clustering~\cite{he2020defending}  &  90.02& 10.09 & 42 \\
Binary weight~\cite{he2020defending} & 89.01 & 10.99 & 89 \\
Model Capacity $\times$ 16~\cite{rakin2021ra} & 93.7 & 10.00 & 49 \\
Weight Reconstruction~\cite{li2020defending}  & 88.79 & 10.00 & 79  \\
RA-BNN~\cite{rakin2021ra}  & 90.18 & 10.00 & 1150 \\ 
\textbf{\emph{DRAM-Locker}}  & \textbf{\emph{91.71}} & \textbf{\emph{ 91.71}} & \textbf{\emph{1150}} \\  \hline
\end{tabular}} \vspace{-1em}
\end{table}


\section{conclusions}
Here, we proposed a general-purpose defense mechanism called  DRAM-Locker that can safeguard DRAM against various adversarial DNN weight attacks affecting data or page tables. DRAM-Locker leverages in-DRAM swapping combined with a lock-table to deliver a high level of protection downgrading the performance of targeted weight attacks to a random attack level. Our defense shows reduced latency, extended defense duration, and no reduction in the accuracy of DNNs when applied to various DNN models compared with existing defense mechanisms.

\bibliographystyle{IEEEtran}
\bibliography{IEEEabrv,./main}\vspace{-2em}
\end{document}